\documentclass[10pt, conference]{IEEEtran}
\IEEEoverridecommandlockouts
\usepackage{cite}
\usepackage{amsmath,amssymb,amsthm,amsfonts}
\usepackage[ruled,vlined]{algorithm2e}
\usepackage{algpseudocode}

\usepackage{graphicx}
\usepackage{textcomp}
\usepackage{xcolor}
\usepackage{microtype}
\usepackage{graphicx}
\usepackage{subfigure}
\usepackage{booktabs} 
\usepackage{paralist}
\usepackage[utf8]{inputenc}

\usepackage[shortlabels]{enumitem}
\usepackage{lipsum, adjustbox}
\usepackage{multirow}
\usepackage{multicol}

\usepackage{pgf-umlsd}
\usepackage{breqn}

\usepackage[font=small, skip=2pt]{caption}
\usepackage{color,soul}

\usepackage{balance}

\usepackage{lipsum, adjustbox}
\usepackage[colorlinks,breaklinks]{hyperref} 
\hypersetup{
	hypertexnames=true, linkcolor=blue, anchorcolor=black,
	citecolor=magenta, urlcolor=alizarin
}

\newsavebox\myboxA
\newsavebox\myboxB
\newlength\mylenA
\newcommand{\overbar}[1]{\mkern 1.5mu\overline{\mkern-1.5mu#1\mkern-1.5mu}\mkern 1.5mu}

\AtBeginDocument{%
  \providecommand\BibTeX{{%
    \normalfont B\kern-0.5em{\scshape i\kern-0.25em b}\kern-0.8em\TeX}}}

\usepackage{color, colortbl}
\definecolor{Gray}{gray}{0.9}
\definecolor{airforceblue}{rgb}{0.36, 0.54, 0.66}
\definecolor{aliceblue}{rgb}{0.94, 0.97, 1.0}
\definecolor{alizarin}{rgb}{0.82, 0.1, 0.26}
\definecolor{amber}{rgb}{1.0, 0.75, 0.0}
\definecolor{amber(sae/ece)}{rgb}{1.0, 0.49, 0.0}
\definecolor{bronze}{rgb}{0.8, 0.5, 0.2}
\definecolor{battleshipgrey}{rgb}{0.52, 0.52, 0.51}
\definecolor{bole}{rgb}{0.47, 0.27, 0.23}
\definecolor{bulgarianrose}{rgb}{0.28, 0.02, 0.03}
\definecolor{cadet}{rgb}{0.33, 0.41, 0.47}
\definecolor{ceil}{rgb}{0.57, 0.63, 0.81}
\definecolor{cerulean}{rgb}{0.0, 0.48, 0.65}
\definecolor{charcoal}{rgb}{0.21, 0.27, 0.31}
\definecolor{coolblack}{rgb}{0.0, 0.18, 0.39}
\definecolor{coolgrey}{rgb}{0.55, 0.57, 0.67}
\definecolor{darkcandyapplered}{rgb}{0.64, 0.0, 0.0}
\definecolor{darkbrown}{rgb}{0.4, 0.26, 0.13}
\definecolor{darkcerulean}{rgb}{0.03, 0.27, 0.49}
\definecolor{darkgray}{rgb}{0.66, 0.66, 0.66}
\definecolor{darkjunglegreen}{rgb}{0.1, 0.14, 0.13}
\definecolor{darktaupe}{rgb}{0.28, 0.24, 0.2}
\definecolor{davy\'sgrey}{rgb}{0.33, 0.33, 0.33}
\definecolor{frenchblue}{rgb}{0.0, 0.45, 0.73}
\definecolor{almond}{rgb}{0.94, 0.87, 0.8}
\definecolor{beaublue}{rgb}{0.74, 0.83, 0.9}
\definecolor{beige}{rgb}{0.96, 0.96, 0.86}
\definecolor{bisque}{rgb}{1.0, 0.89, 0.77}
\definecolor{black}{rgb}{0.0, 0.0, 0.0}
\definecolor{fluorescentorange}{rgb}{1.0, 0.75, 0.0}
\definecolor{ghostwhite}{rgb}{0.97, 0.97, 1.0}
\definecolor{antiquewhite}{rgb}{0.98, 0.92, 0.84}
\definecolor{LightCyan}{rgb}{0.88,1,1}

\theoremstyle{plain}

\theoremstyle{definition}

\theoremstyle{remark}

\usepackage[textsize=tiny]{todonotes}

\def\BibTeX{{\rm B\kern-.05em{\sc i\kern-.025em b}\kern-.08em
    T\kern-.1667em\lower.7ex\hbox{E}\kern-.125emX}}
\begin{document}

\title{\huge Love or Hate? Share or Split? Privacy-Preserving Training Using Split Learning and Homomorphic Encryption
}

\author{\IEEEauthorblockN{1\textsuperscript{st} Tanveer Khan}
\IEEEauthorblockA{
\textit{Tampere University}\\
Tampere, Finland \\
tanveer.khan@tuni.fi}
\and
\IEEEauthorblockN{2\textsuperscript{nd} Khoa Nguyen}
\IEEEauthorblockA{
\textit{Tampere University}\\
Tampere, Finland \\
khoa.nguyen@tuni.fi}
\and
\IEEEauthorblockN{3\textsuperscript{rd} Antonis Michalas}
\IEEEauthorblockA{
\textit{Tampere University, Finland,}\\
RISE Research Institutes of Sweden \\
antonios.michalas@tuni.fi}
\and
\IEEEauthorblockN{4\textsuperscript{th} Alexandros Bakas}
\IEEEauthorblockA{
\textit{Tampere University and Nokia Bell Labs}\\
Tampere, Finland \\
alexandros.bakas@tuni.fi}
}

\maketitle

\begin{abstract}
Split learning (SL) is a new collaborative learning technique that allows participants, e.g.\ a client and a server, to train machine learning models without the client sharing raw data. In this setting, the client initially applies its part of the machine learning model on the raw data to generate activation maps and then sends them to the server to continue the training process. Previous works in the field demonstrated that reconstructing activation maps could result in privacy leakage of client data. In addition to that, existing mitigation techniques that overcome the privacy leakage of SL prove to be significantly worse in terms of accuracy. In this paper, we improve upon previous works by constructing a protocol based on U-shaped SL that can operate on homomorphically encrypted data. More precisely, in our approach, the client applies homomorphic encryption on the activation maps before sending them to the server, thus protecting user privacy. This is an important improvement that reduces privacy leakage in comparison to other SL-based works. Finally, our results show that, with the optimum set of parameters, training with HE data in the U-shaped SL setting only reduces accuracy by 2.65\% compared to training on plaintext. In addition, raw training data privacy is preserved.
\end{abstract}

\begin{IEEEkeywords}
Activation Maps, Homomorphic Encryption, Machine Learning, Privacy, Split Learning
\end{IEEEkeywords}

\section{Introduction}



SL~\cite{gupta2018distributed} is a collaborative learning approach that divides a Machine Learning (ML)  into two parts: the client-side and the server-side. It facilitates training the Deep Neural Networks (DNN) among multiple data sources, while mitigating the need to directly share raw labeled data with collaboration parties. 
The advantages of SL are multifold: \begin{inparaenum}[\it (i)] \item it allows multiple parties to collaboratively train a neural network, \item it allows users to train ML models without sharing their raw data with a server running part of a DNN model, thus preserving user privacy, \item it protects both the client and the server from revealing their parts of the model, and \item it reduces the client's computational overhead by not running the entire model (i.e.\ utilizing a smaller number of layers)~\cite{vepakomma2019split}.\end{inparaenum}

Though SL offers an extra layer of privacy protection by definition, there are no works exploring how it is combined with popular techniques that promise to preserve user privacy (e.g.\ encryption). In~\cite{abuadbba2020can}, the authors studied whether SL can handle sensitive time-series data and demonstrated that SL alone is \textit{insufficient} when performing privacy-preserving training for 1-dimensional (1D) CNN models. More precisely, the authors showed raw data can be reconstructed from the activation maps of the intermediate split layer. The authors also employed two mitigation techniques, adding hidden layers and applying differential privacy to reduce privacy leakage. However, based on the results, none of these techniques can effectively reduce privacy leakage from all channels of the SL activation. Furthermore, both these techniques result in significantly reducing the joint model's accuracy.

In this paper, we focus on training an ML model in a privacy-preserving manner, where a client and a server collaborate to train the model. More specifically, we construct a model that uses Homomorphic Encryption (HE)~\cite{cheon2017homomorphic} to mitigate privacy leakage in SL. In our model, the client first encrypts the activation maps and then sends the encrypted activation maps to the server. The encrypted activation maps do \textit{not} reveal anything about the raw data (i.e.\ it is \textit{not} possible to reconstruct the original raw data from the encrypted activation maps).


\medskip 
\noindent \textbf{\textit{Contributions:}} \enskip
The main contributions of this paper are:  
\begin{enumerate}[\bfseries C1.]
    \item We construct a U-shaped split 1D CNN model and experiment using plaintext activation maps sent from the client to the server. Through the U-shaped 1D CNN model, clients do \textit{not} need to share the ground truth labels with the server -- this is an important improvement that reduces privacy leakage compared to~\cite{abuadbba2020can}.
    \item We constructed the HE version of the U-shaped SL technique. In the encrypted U-shaped SL model, the client encrypts the activation map using HE and sends it to the server. The advantage of HE encrypted U-shaped SL over the plaintext U-shaped SL is that server performs computation over encrypted activation maps.
    \item To assess the applicability of our framework, we performed experiments on the PTB-XL heartbeat datasets~\cite{wagner2020ptb}, currently being the largest open-source electrocardiography dataset to our knowledge. 
    We experimented with activation maps of lengths 256 for both plaintext and homomorphically encrypted activation maps and we have measured the model's performance by considering training duration, test accuracy, and communication cost. 
\end{enumerate}

\medskip 

\noindent \textbf{\textit{Organization:}} \enskip
	The rest of the paper is organized as follows: In \autoref{sec:preliminaries}, we provide necessary background for 1D CNN, HE and SL. 
	In \autoref{sec:relatedwork}, we present works in the area of SL, followed by
	design and implementation of the split 1D CNN training protocols in \autoref{sec:slProtocol}, 
 extensive experimental results in \autoref{sec:performance}
and conclude the paper in~\autoref{sec:Conclusion}.

\section{Preliminaries}
\label{sec:preliminaries}
\subsection{Convolutional Neural Network}
CNN is a special type of neural network using convolution layers to extract features from data~\cite{lecun1998CNN}. 
In this work, we employ a 1D CNN~\cite{li2017classification,abuadbba2020can} as a feature detector and classifier for a heartbeat datasets, namely PTB-XL~\cite{wagner2020ptb}. The employed 1D CNN consists of the following layers stacked on top of each other, namely: Conv1D, Leaky ReLU, max pooling and a single fully connected and a softmax layer (see \autoref{fig:u-shapedSL}).

\subsection{Homomorphic Encryption}
HE is an emerging cryptographic technique for computations on encrypted data. HE schemes are divided into three main categories according to their functionality: Partial HE~\cite{paillier1999public}, leveled (or somewhat) HE~\cite{cheon2017homomorphic}, and fully HE~\cite{gentry2009fully}. Each scheme has its own benefits and disadvantages. In this work, we use the CKKS Leveled HE scheme~\cite{cheon2017homomorphic}. CKKS allows users to do additions and a limited number of multiplications on vectors of complex values (and hence, real values too). 
The most important parameters of the CKKS scheme are Polynomial Modulus $\mathcal{P}$, Coefficient Modulus $\mathcal{C}$ and Scaling Factor $\Delta$.

\subsection{Split Learning}
\label{subsubsec:splitlear}
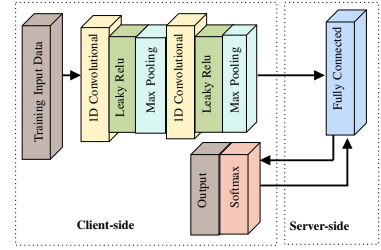
\begin{figure}
 \centering
 \begin{adjustbox}{width=0.27\textwidth}
		\tikzset{every picture/.style={line width=0.75pt}} 
\begin{tikzpicture}[x=0.75pt,y=0.75pt,yscale=-1,xscale=1]
\draw  [fill={rgb, 255:red, 255; green, 244; blue, 199 }  ,fill opacity=1 ] (167,155.9) -- (179.9,143) -- (210,143) -- (210,265.1) -- (197.1,278) -- (167,278) -- cycle ; \draw   (210,143) -- (197.1,155.9) -- (167,155.9) ; \draw   (197.1,155.9) -- (197.1,278) ;
\draw  [fill={rgb, 255:red, 200; green, 218; blue, 164 }  ,fill opacity=1 ] (197,165.6) -- (209.6,153) -- (239,153) -- (239,255.5) -- (226.4,268.1) -- (197,268.1) -- cycle ; \draw   (239,153) -- (226.4,165.6) -- (197,165.6) ; \draw   (226.4,165.6) -- (226.4,268.1) ;
\draw  [fill={rgb, 255:red, 218; green, 246; blue, 242 }  ,fill opacity=1 ] (225.2,166.8) -- (239,153) -- (271.2,153) -- (271.2,255.2) -- (257.4,269) -- (225.2,269) -- cycle ; \draw   (271.2,153) -- (257.4,166.8) -- (225.2,166.8) ; \draw   (257.4,166.8) -- (257.4,269) ;
\draw  [fill={rgb, 255:red, 197; green, 181; blue, 175 }  ,fill opacity=1 ] (104,153.9) -- (116.9,141) -- (147,141) -- (147,284.1) -- (134.1,297) -- (104,297) -- cycle ; \draw   (147,141) -- (134.1,153.9) -- (104,153.9) ; \draw   (134.1,153.9) -- (134.1,297) ;
\draw  [fill={rgb, 255:red, 255; green, 244; blue, 199 }  ,fill opacity=1 ] (258.4,153.9) -- (271.3,141) -- (301.4,141) -- (301.4,262.1) -- (288.5,275) -- (258.4,275) -- cycle ; \draw   (301.4,141) -- (288.5,153.9) -- (258.4,153.9) ; \draw   (288.5,153.9) -- (288.5,275) ;
\draw  [fill={rgb, 255:red, 200; green, 218; blue, 164 }  ,fill opacity=1 ] (289,165.6) -- (301.6,153) -- (331,153) -- (331,255.4) -- (318.4,268) -- (289,268) -- cycle ; \draw   (331,153) -- (318.4,165.6) -- (289,165.6) ; \draw   (318.4,165.6) -- (318.4,268) ;
\draw  [fill={rgb, 255:red, 218; green, 246; blue, 242 }  ,fill opacity=1 ] (318.4,163.98) -- (329.38,153) -- (355,153) -- (355,257.02) -- (344.02,268) -- (318.4,268) -- cycle ; \draw   (355,153) -- (344.02,163.98) -- (318.4,163.98) ; \draw   (344.02,163.98) -- (344.02,268) ;
\draw  [fill={rgb, 255:red, 195; green, 220; blue, 252 }  ,fill opacity=1 ] (429.4,149.9) -- (442.3,137) -- (472.4,137) -- (472.4,258.1) -- (459.5,271) -- (429.4,271) -- cycle ; \draw   (472.4,137) -- (459.5,149.9) -- (429.4,149.9) ; \draw   (459.5,149.9) -- (459.5,271) ;
\draw  [fill={rgb, 255:red, 197; green, 181; blue, 175 }  ,fill opacity=1 ] (284.5,291.9) -- (297.4,279) -- (327.5,279) -- (327.5,365.16) -- (314.6,378.06) -- (284.5,378.06) -- cycle ; \draw   (327.5,279) -- (314.6,291.9) -- (284.5,291.9) ; \draw   (314.6,291.9) -- (314.6,378.06) ;
\draw  [fill={rgb, 255:red, 246; green, 200; blue, 185 }  ,fill opacity=1 ] (315.6,291.9) -- (328.5,279) -- (358.6,279) -- (358.6,364.17) -- (345.7,377.07) -- (315.6,377.07) -- cycle ; \draw   (358.6,279) -- (345.7,291.9) -- (315.6,291.9) ; \draw   (345.7,291.9) -- (345.7,377.07) ;
\draw [line width=1.5]    (147,207) -- (163,207) ;
\draw [shift={(167,207)}, rotate = 180] [fill={rgb, 255:red, 0; green, 0; blue, 0 }  ][line width=0.08]  [draw opacity=0] (11.61,-5.58) -- (0,0) -- (11.61,5.58) -- cycle    ;
\draw [line width=1.5]    (356,207) -- (424,207) ;
\draw [shift={(428,207)}, rotate = 180] [fill={rgb, 255:red, 0; green, 0; blue, 0 }  ][line width=0.08]  [draw opacity=0] (11.61,-5.58) -- (0,0) -- (11.61,5.58) -- cycle    ;
\draw [line width=1.5]    (437,298) -- (362,298) ;
\draw [shift={(358,298)}, rotate = 360] [fill={rgb, 255:red, 0; green, 0; blue, 0 }  ][line width=0.08]  [draw opacity=0] (11.61,-5.58) -- (0,0) -- (11.61,5.58) -- cycle    ;
\draw  [dash pattern={on 0.84pt off 2.51pt}] (97,129) -- (376.5,129) -- (376.5,389) -- (97,389) -- cycle ;
\draw  [dash pattern={on 0.84pt off 2.51pt}] (385,129) -- (485.5,129) -- (485.5,388) -- (385,388) -- cycle ;
\draw [line width=1.5]    (437,271) -- (437,298) ;
\draw [line width=1.5]    (359,325) -- (453,325) ;
\draw [line width=1.5]    (452,325) -- (452,278) ;
\draw [shift={(452,274)}, rotate = 90] [fill={rgb, 255:red, 0; green, 0; blue, 0 }  ][line width=0.08]  [draw opacity=0] (11.61,-5.58) -- (0,0) -- (11.61,5.58) -- cycle    ;
\draw (174.42,260.07) node [anchor=north west][inner sep=0.75pt]  [rotate=-269.64] [align=left] {1D Convolutional};
\draw (265.42,260.07) node [anchor=north west][inner sep=0.75pt]  [rotate=-269.64] [align=left] {1D Convolutional};
\draw (202.42,253.07) node [anchor=north west][inner sep=0.75pt]  [rotate=-269.64] [align=left] {Leaky Relu};
\draw (295.42,249.07) node [anchor=north west][inner sep=0.75pt]  [rotate=-269.64] [align=left] {Leaky Relu};
\draw (235.42,249.07) node [anchor=north west][inner sep=0.75pt]  [rotate=-269.64] [align=left] {Max Pooling};
\draw (325.42,249.07) node [anchor=north west][inner sep=0.75pt]  [rotate=-269.64] [align=left] {Max Pooling};
\draw (161,361) node [anchor=north west][inner sep=0.75pt]   [align=left] {\textbf{Client-side}};
\draw (389,362) node [anchor=north west][inner sep=0.75pt]   [align=left] {\textbf{Server-side}};
\draw (325.42,350.83) node [anchor=north west][inner sep=0.75pt]  [rotate=-269.64] [align=left] {Softmax};
\draw (290.42,350.28) node [anchor=north west][inner sep=0.75pt]  [rotate=-269.64] [align=left] {Output};
\draw (115.42,280.07) node [anchor=north west][inner sep=0.75pt]  [rotate=-269.64] [align=left] {Training Input Data};
\draw (435.42,246.07) node [anchor=north west][inner sep=0.75pt]  [rotate=-269.64] [align=left] {Fully Connected};
\end{tikzpicture}
  \end{adjustbox}
		\caption{U-shaped Split-Learning}
		\label{fig:u-shapedSL}
	\end{figure}

Although the configuration of SL could potentially take many forms, widely known configurations are simple vanilla split and U-shaped SL. In a simple two-party vanilla split, a client and a server collaborate to train the split model with no access to each other's parts. More specifically, the layers of a DNN model are split into two parts (A and B) and distributed to the client and the server. The client owning the data, uses forward propagation to train part A (comprising the first few layers) and sends the activated output from the split layer (part A's final layer) to the server. The server performs the forward training on the outputs activated from the client using part B. Furthermore, the server performs the backward propagation on part B, only returning to the client the gradients of the split layer's active outputs (part B's first layer) to complete the backward propagation on part A. This process is repeated until the model converges. Although the client and server do not share any raw input data, this configuration requires label sharing. The sharing of labels can be eliminated by using a U-shaped configuration. The U-shaped SL configuration is different from the simple vanilla setup, as it does not require of clients to share labels~\cite{vepakomma2019split}. On the server-side, the network is wrapped at the end layer, and the outputs are sent back to client entities (see \autoref{fig:u-shapedSL}). Clients generate gradients from the end layers and utilize them for backward propagation without revealing the corresponding labels.

\section{Related Work}
\label{sec:relatedwork}

Initially, it was believed that SL is a promising approach in terms of client raw data protection, and for doing PPML~\cite{khan2021blind} as it does \textit{not} share raw data and also has the benefit of \textit{not} disclosing the model's architecture and weights.
For example, Gupta and Raskar~\cite{gupta2018distributed} predicted that reconstructing raw data on the client-side, while using SL would be difficult. In addition, the authors of~\cite{vepakomma2019split} employed the SL model to the healthcare applications to protect the users' personal data. Vepakomma \textit{et al.} found that SL outperforms other collaborative learning techniques like federated learning~\cite{yang2019federated} in terms of accuracy~\cite{vepakomma2019split}.

However, SL provides data privacy on the grounds that only activation maps are shared between the parties. Different studies showed the possibility of privacy leakage in SL. In~\cite{vepakomma2019reducing}, the authors analyzed the privacy leakage of SL and found a considerable leakage from the split layer in the 2D~CNN model. Furthermore, the authors mentioned that it is possible to reduce the distance correlation (a measure of dependence) between the split layer and raw data by slightly scaling the weights of all layers before the split. This works well in models with a large number of hidden layers before the split. 

The work of Abuadbba \textit{et al.}~\cite{abuadbba2020can} is the first study exploring whether SL can deal with time-series data. 
According to the results, SL can be applied to a model without the model accuracy degradation. 
The authors also showed that when SL is directly adopted into 1D CNN models for time series data could result in significant privacy leakage. Two mitigation techniques were employed to limit the potential privacy leakage in SL: \begin{inparaenum}[\it (i)] \item increasing the number of layers before the split on the client-side and \item applying differential privacy to the split layer activation before sending the activation map to the server\end{inparaenum}. However, both techniques suffer from a loss of model accuracy, particularly when differential privacy is used. 
	
In~\cite{abuadbba2020can}, during the forward propagation, the client sends the activation map in plaintext to the server, while the server can easily reconstruct the original raw data from the activated vector of the split layer leading to clear privacy leakage. In our work, we constructed a training protocol, where, instead of sending plaintext activation maps, the client first conducts an encryption using HE and then sends said maps to the server. In this way, the server is unable to reconstruct the original raw data, but can still perform a computation on the encrypted activation maps, enabling the the training process.

\smallskip

\section{Split Model Training Protocols}
\label{sec:slProtocol}
In this section, we first present the protocol for training the U-shaped split 1D CNN on plaintext activation maps, followed by the protocol for training the U-shaped split 1D CNN on encrypted activation maps.

\subsection{Actors in the U-Split Learning Model}
\label{subsec:slActors}
We have two parties: client and a server. Each party plays a specific role and has access to certain parameters.   We construct the U-shaped split 1D CNN in such a way that the first few as well as the last layer are on the client-side, while the remaining layers are on the server-side, as demonstrated in \autoref{fig:u-shapedSL}. The client and server collaborate to train the split model by sharing the activation maps and gradients. To elaborate further, we describe their respective roles and access as follows:
\begin{itemize}
    \item Client: In the plaintext version, the client holds two Conv1D layers and can access their weights and biases in plaintext. However, other layers such as Max Pooling, Leaky ReLU, Softmax do not have associated weights and biases. Apart from these, in the HE encrypted version, the client is also responsible for generating the context for HE and has access to all
    context parameters including Polynomial modulus $\mathcal{P}$, Coefficient modulus $\mathcal{C}$, Scaling factor $\Delta$, Public key $\mathsf{pk}$ and Secret key $\mathsf{sk}$. Note that for both training on plaintext and encrypted activation maps, the raw data examples $\mathbf{x}$'s and their corresponding labels $\mathbf{y}$'s reside on the client side and are never sent to the server during the training process.
    \item Server: The computation performed on the server-side is limited to only one linear layer. The reason for having only one linear layer on the server-side is due to computational constraints when training on HE encrypted data. Hence, the server 
    can exclusively access 
    the weights and biases of this linear layer. Regarding the HE context parameters, the server has access to $\mathcal{P}$, $\mathcal{C}$, $\Delta$, and $\mathsf{pk}$ shared by the client, with the exception of 
    the $\mathsf{sk}$. 
    Not holding the secret key, the server cannot decrypt the HE encrypted activation maps sent from the client. The hyperparameters shared between the client and the server include the learning rate ($\eta$), batch size ($n$), number of batches to be trained ($N$), and the number of training epochs ($E$).
\end{itemize}

\subsection{Plaintext Activation Maps}
\label{subsection: unencryptedactivation}
We have used ~\autoref{alg:client} and~\autoref{alg:server} to train the U-shaped split 1D CNN model. First, the client and server start the socket initialization process and synchronize the hyperparameters $\eta, n, N, E$. They also initialize the weights of their layers using the randomly initialized weights $\Phi$.

During the forward propagation phase, the client forward-propagates the input $\mathbf{x}$ until the $l^{th}$ layer and sends the activation $\mathbf{a}^{(l)}$ to the server. The server continues 
to forward propagate and sends the output $\mathbf{a}^{(L)}$ to the client. Next, the client applies the Softmax function on $\mathbf{a}^{(L)}$ to get $\mathbf{\hat{y}}$ and calculates the error $J$ using the loss function $J = \mathcal{L}(\mathbf{\hat{y}}, \mathbf{y})$.

The client starts the backward propagation by calculating and sending the gradient of the error w.r.t $\mathbf{a}^{(L)}$, i.e. $\frac{\partial J}{\partial \mathbf{a}^{(L)}}$, to the server. The server continues the backward propagation, calculates $\frac{\partial J}{\partial \mathbf{a}^{(l)}}$ and sends $\frac{\partial J}{\partial \mathbf{a}^{(l)}}$ to the client. After receiving the gradients $\frac{\partial J}{\partial \mathbf{a}^{(l)}}$ from the server, the backward propagation continues to the first hidden layer on the client-side. Note that the exchange of information between client and server in these algorithms takes place in plaintext. As can be seen in~\autoref{alg:client}, the client sends the activation maps $\mathbf{a}^{(l)}$ to the server in plaintext and receives the output of the linear layer $\mathbf{a}^{(L)}$ from the server in plaintext. 	
The same applies on the server side: receiving $\mathbf{a}^{(l)}$ and sending $\mathbf{a}^{(L)}$ in the plaintext as can be seen in~\autoref{alg:server}. Sharif \textit{et al.}~\cite{abuadbba2020can} showed that the exchange of plaintext activation maps between client and server using SL reveals important information regarding the client's raw sequential data. Later, in~\autoref{subsec:VisualInvert} we show in detail how passing the forward activation maps from the client to the server in the plaintext will result in information leakage. To mitigate this privacy leakage, we propose the protocol, where the client encrypts the activation maps before sending them to the server, as described in~\autoref{subsubsection: encryptedactivation}.

\begin{algorithm}[!h]
\scriptsize 
\SetAlgoLined
 \textbf{Initialization:}\\
 $s\leftarrow$ socket initialized with port and address\;
 \textit{s.connect}\\
 $\eta, n, N, E \leftarrow s.synchronize()$\\
 $ \{\boldsymbol{w}^{( i)}, \boldsymbol{b}^{( i)}\}_{\forall i\in \{0..l\}} \ \leftarrow \ initialize\ using\ \Phi $\\
 $\{\mathbf{z}^{( i)}\}_{\forall i\in \{0..l\}} ,\{\mathbf{a}^{( i)}\}_{\forall i\in \{0..l\}}\leftarrow \emptyset \ $\\
 $ \left\{\frac{\partial J}{\partial \mathbf{z}^{( i)}}\right\}_{\forall i\in \{0..l\}} ,\left\{\frac{\partial J}{\partial \mathbf{a}^{( i)}}\right\}_{\forall i\in \{0..l\}}\leftarrow \emptyset \ $\\
 \For{$\displaystyle e \ \in \ E $}{
 	\For{$\displaystyle \text{each} \ \text{batch}\ ( \mathbf{x},\ \mathbf{y}) \ \text{generated\ from}\ D\ $}{
 	      $\displaystyle  \mathbf{Forward\ propagation:}$\\
          $\displaystyle \ \ \ \ O.zero\_grad()  $\\
 	$\displaystyle \ \ \ \ \mathbf{a}^{0} \ \ \leftarrow \mathbf{x}$ \\
 	\For{$i \leftarrow 1$ to $l$}{
 	$\displaystyle \mathbf{z}^{( i)} \ \leftarrow \ f^{( i)}\left( \mathbf{a}^{( i-1)}\right)$\\
 	$\displaystyle \mathbf{a}^{( i)} \ \leftarrow \ g^{( i)}\left( \mathbf{z}^{( i)}\right)$\\}
 	$\displaystyle \ \ \ \  s.send\ ( \mathbf{a}^{(l)})$\\
 	$\displaystyle \ \ \ \ s.receive\ ( \mathbf{a}^{(L)})$\\
 	$\displaystyle \ \ \ \  \hat{y} \ \leftarrow \ Softmax\left(\mathbf{a}^{( L)}\right)$\\
 	$\displaystyle \ \ \ \ J \leftarrow \mathcal{L} (\hat{\mathbf{y}}, \mathbf{y})$\\
 	$\displaystyle \mathbf{Backward\ propagation:}$\\
 	$\displaystyle \ \ \ \ \text{Compute}\left\{\frac{\partial J}{\partial \hat{\mathbf{y}}}\ \&\ \frac{\partial J}{\partial \mathbf{a}^{(L)}}\right\}$\\
 	$\displaystyle \ \ \ \ s.send\ \left( \frac{\partial J}{\partial \mathbf{a}^{(L)}} \right)$\\
 	$\displaystyle \ \ \ \ s.receive\ \left( \frac{\partial J}{\partial \mathbf{a}^{( l)}} \right)$\\
 	\For{$i\leftarrow 1$ to $l$}{$\text{Compute}\ \left\{ \frac{\partial J}{\partial \boldsymbol{w}^{( i)}}, \ \frac{\partial J}{\partial \boldsymbol{b}^{( i)}} \right\}$\\
 	$\displaystyle\ \ \ \ \ \ \ \ \text{Update}\ \boldsymbol{w}^{( i)},\ \boldsymbol{b}^{( i)}$
 	}
 	}
 }
 \caption{\textbf{Client Side}}
 \label{alg:client}
\end{algorithm}	

\subsection{Encrypted Activation Maps}
\label{subsubsection: encryptedactivation}
The protocol for training the U-shaped 1D CNN with a homomorphically encrypted activation map consists of four phases: initialization, forward propagation, classification, and backward propagation. The initialization phase only takes place once at the beginning of the procedure, whereas the other phases continue until the model iterates through all epochs.
Each phase is described in detail in the following subsections.

\begin{algorithm}
\scriptsize 
\SetAlgoLined
 \textbf{Initialization:}\\
 $s\leftarrow$ socket initialized with port and address\;
 \textit{s.connect}\\
 $\eta, n, N, E \leftarrow s.synchronize()$\\
 $ \{\boldsymbol{w}^{( i)}, \boldsymbol{b}^{( i)}\}_{\forall i\in \{0..l\}} \ \leftarrow \ initialize\ using\ \Phi $\\
 $\displaystyle \{\mathbf{z}^{( i)}\}_{\forall i\in \{l+1..L\}} \leftarrow \emptyset \ $\\
 $\displaystyle  \left\{\frac{\partial J}{\partial \mathbf{z}^{( i)}}\right\}_{\forall i\in \{l+1..L\}} \leftarrow \emptyset \ $\\
 \For{$\displaystyle e \ \in \ E $}{
 	\For{$\displaystyle i \leftarrow 1 \ \mathbf{to} \ N \ $}{
 	$\displaystyle \mathbf{Forward\ propagation:}$\\
 	$\displaystyle \ \ \ \ O.zero\_grad()  $\\
 	$\displaystyle \ \ \ \ s.receive\ (\mathbf{a}^{(l)}) \ \ $ \\
$\displaystyle \ \ \ \ \mathbf{a}^{(L)} \ \leftarrow \ f^{( i)}\left( \mathbf{a}^{(l)}\right)$\\
$\displaystyle \ \ \ \ s.send\left( \mathbf{a}^{(L)}\right)$\\
$\displaystyle \mathbf{Backward\ propagation:}$\\
$\displaystyle \ \ \ \ s.receive \ \left( \frac{\partial J}{\partial \mathbf{a}^{(L)}}\right)$\\
$\displaystyle \ \ \ \ \text{Compute}\ \left\{ \frac{\partial J}{\partial \boldsymbol{w}^{(L)}}, \ \frac{\partial J}{\partial \boldsymbol{b}^{(L)}} \right\}$\\
$\displaystyle \ \ \ \ \text{Update}\ \boldsymbol{w}^{( L)},\ \boldsymbol{b}^{(L)}  $\\
$\displaystyle \ \ \ \ \text{Compute}\ \frac{\partial J}{\partial \mathbf{a}^{( l)}} $\\
$\displaystyle \ \ \ \ s.send \left( \frac{\partial J}{\partial \mathbf{a}^{( l)}} \right)$\\
 		}
 }
 \caption{\textbf{Server Side}}
 \label{alg:server}
\end{algorithm}

\paragraph*{Initialization} The initialization phase consists of socket initialization, context generation, and random weight loading. The client first establishes a socket connection to the server and synchronizes the four hyperparameters $\eta ,\ n,\ N, E $ with the server, shown in~\autoref{alg:clientHE} and~\autoref{alg:serverHE}. These parameters must be synchronized on both sides to be trained in the same way. Also, the weights on the client and server are initialized with the same set of corresponding weights in the local model to accurately assess and compare the influence of SL on performance. On both the client and the server sides, $\boldsymbol{w}^{(i)}$ are initialized using corresponding parts of $\Phi$. The activation map at layer $i$ ($\mathbf{a}^{(i)}$), output tensor of Conv1D layer ($\mathbf{z}^{(i)}$), and the gradients are initially set to zero. In this phase, the context generated is a specific object that holds $\mathsf{pk}$ and $\mathsf{sk}$ of the HE scheme as well as certain other parameters like $\mathcal{P}$, $\mathcal{C}$ and $\Delta$.

Further information on the HE parameters and how to choose the best-suited parameters can be found in the~\href{https://bit.ly/3KY8ByN}{TenSEAL's benchmarks tutorial}. As shown in~\autoref{alg:clientHE} and~\autoref{alg:serverHE}, the context is either public ($\mathsf{ctx_{pub}}$) or private ($\mathsf{ctx_{pri}}$) depending on whether it holds the secret key $\mathsf{sk}$. Both the $\mathsf{ctx_{pub}}$ and $\mathsf{ctx_{pri}}$ have the same parameters, though $\mathsf{ctx_{pri}}$ holds a $\mathsf{sk}$ and $\mathsf{ctx_{pub}}$ does not. The server does not have access to the $\mathsf{sk}$ as the client only shares the $\mathsf{ctx_{pub}}$ with the server. After completing the initialization phase, both the client and server proceed to the forward and backward propagation phases.

\begin{algorithm}[!h]
\scriptsize 
\SetAlgoLined
 \textbf{Context Initialization:}\\
 $\displaystyle \ \ \ \ \ \mathsf{ctx_{pri}},\ \leftarrow \ \mathcal{P}, \ \mathcal{C}, \ \Delta, \  \mathsf{pk}, \ \mathsf{sk}$\\
 $\displaystyle \ \ \ \ \ \mathsf{ctx_{pub}},\ \leftarrow \ \mathcal{P}, \ \mathcal{C}, \ \Delta, \  \mathsf{pk}$\\
			 $\displaystyle \ \ \ \ \ s.send( \mathsf{ctx_{pub}})$\\
	\For{$\displaystyle e \ \text{in} \ E $}{
	\For{$\displaystyle \text{ each} \ \text{batch}\ ( \mathbf{x},\ \mathbf{y}) \ \text{generated\ from}\ \mathbf{D}\ $}{
	$\displaystyle  \mathbf{Forward\ propagation:}$\\
	$\displaystyle \ \ \ \ O.zero\_grad()  $\\
	$\displaystyle \ \ \ \ \mathbf{a}^{0}\ \ \leftarrow \mathbf{x}$ \\
	\For{$i\ \leftarrow \ 1\ \mathbf{to} \ l$}{
	$\displaystyle \ \ \ \ \ \ \ \ \mathbf{z}^{( i)} \ \leftarrow \ f^{( i)}\left( \mathbf{a}^{( i-1)}\right)$\\
	$\displaystyle \ \ \ \ \ \ \ \ \mathbf{a}^{i} \ \leftarrow \ g^{( i)}\left( \mathbf{z}^{( i)}\right)$}
	$\displaystyle \ \ \ \ \overbar{\mathbf{a}^{(l)}} \ \leftarrow \ \mathsf{HE.Enc}\left(\mathsf{pk}, \mathbf{a}^{(l)}\right)$\\
	$\displaystyle \ \ \ \ s.send \ \overbar{(\mathbf{a}^{(l)})}$\\
	$\displaystyle \ \ \ \  s.receive\ ( \overbar{\mathbf{a}^{(L)})}$\\
	$\displaystyle  \ \ \ \  \mathbf{a}^{( L)} \ \leftarrow \ \mathsf{HE.Dec}\left(\mathsf{sk}, \overbar{\mathbf{a}^{( L)}}\right)$\\
	$\displaystyle  \ \ \ \  \hat{\mathbf{y}} \ \leftarrow \ Softmax\left(\mathbf{a}^{( L)}\right)$\\
	$\displaystyle \ \ \ \  \mathbf{J} \leftarrow \mathcal{L} (\hat{\mathbf{y}}, \mathbf{y})$\\
	$\displaystyle \mathbf{Backward\ propagation:}$\\
	$\displaystyle \ \ \ \ \text{Compute}\left\{\frac{\partial J}{\partial \hat{\mathbf{y}}} \& \frac{\partial J}{\partial \mathbf{a}^{(L)} } \& \frac{\partial J}{\partial \boldsymbol{w}^{(L)}} \right\}$\\
	$\displaystyle \ \ \ \ s.send\left(\frac{\partial J}{\partial \mathbf{a}^{(L)} } \& \frac{\partial J}{\partial \boldsymbol{w}^{(L)}}\right)$\\
	$\displaystyle \ \ \ \ s.receive\left( \frac{\partial J}{\partial \mathbf{a}^{(l)}} \right)$\\
	\For{$i\leftarrow l\ \text{down to} \ 1$}{
	$\displaystyle \ \ \ \ \ \ \ \ \text{Compute} \left\{ \frac{\partial J}{\partial \boldsymbol{w}^{(i)}}, \ \frac{\partial J}{\partial \boldsymbol{b}^{(i)}} \right\}$\\
	$\displaystyle \ \ \ \ \ \ \ \ \text{Update}\ \boldsymbol{w}^{( i)},\ \boldsymbol{b}^{(i)} $}
	}	
	}
 \caption{\textbf{Client Side}}
 \label{alg:clientHE}
\end{algorithm}

\paragraph*{Forward propagation} The forward propagation starts on the client side. The client first zeroes out the gradients for the batch of data $(\mathbf{x}, \mathbf{y})$. He then begins calculating $\mathbf{a}^{(l)}$  from $\mathbf{x}$, ~\autoref{alg:clientHE} where each $f^{(i)}$ is a Conv1D layer.

The~\href{https://pytorch.org/docs/stable/generated/torch.nn.Conv1d.html}{Conv1D} layer can be described as following: given a 1D input signal that contains $C$ channels, where each channel $\mathbf{x}_{(i)}$ is a 1D array ($i\in \{1,\ldots,C\}$), a Conv1D layer produces an output that contains $C'$ channels. The $j^{th}$ output channel $\mathbf{y}_{(j)}$, where $j\in \{1,\ldots,C'\}$, can be described as
\begin{equation}\label{eq:1dconvOp}
	\mathbf{y}_{(j)} = \boldsymbol{b}_{(j)}  + \sum_{i=1}^{C} \boldsymbol{w}_{(i)} \star \mathbf{x}_{(i)},
\end{equation}
where $\boldsymbol{w}_{(i)}, i\in \{1,\ldots,C\}$ are the weights, $\boldsymbol{b}_{(j)}$ are the biases of the Conv1D layer, and $\star$ is the 1D cross-correlation operation. The $\star$ operation can be described as
\begin{equation}
	\mathbf{z}(i) = (\boldsymbol{w} \star \mathbf{x}) (i) = \sum_{j=0}^{m-1}\boldsymbol{w}(j) \cdot \mathbf{x}(i+j), 
\end{equation}
where $\mathbf{z}(i)$ denotes the $i^{th}$ element of the output vector $\mathbf{z}$, and $i$ starts at 0. Here, the size of the 1D weighted kernel is $m$. 

In~\autoref{alg:clientHE}, $g^{(i)}$ can be seen as the combination of Max Pooling and Leaky ReLU functions. The final output activation maps of the $l^{th}$ layer from the client is $\mathbf{a}^{(l)}$. The client then homomorphically encrypts $\mathbf{a}^{(l)}$ and sends the encrypted activation maps $\overbar{\mathbf{a}^{(l)}}$ to the server. In~\autoref{alg:serverHE}, the server receives $\overbar{\mathbf{a}^{(l)}}$ and then performs forward propagation, which is a linear layer evaluated on HE encrypted data $\overbar{\mathbf{a}^{(l)}}$ as
\begin{equation}
	\label{eq:serverHELinear}
	\overbar{\mathbf{a}^{(L)}} = \overbar{\mathbf{a}^{(l)}} \boldsymbol{w}^{(L)} + \boldsymbol{b}^{(L)} .
\end{equation}
After that, the server sends $\overbar{\mathbf{a}^{(L)}}$ to the client~(\autoref{alg:serverHE}). Upon reception, the client decrypts $\overbar{\mathbf{a}^{(L)}}$ to get $\mathbf{a}^{(L)}$, performs Softmax on $\mathbf{a}^{(L)}$ to produce the predicted output $\mathbf{\hat{y}}$ and calculate the loss $J$, as can be seen in~\autoref{alg:clientHE}. 
Having finished the forward propagation 
we may move on to the backward propagation part of the protocol.

\paragraph*{Backward propagation} After calculating $J$, the client starts the backward propagation by initially computing 
$\frac{\partial J}{\partial \hat{\mathbf{y}}}$ and then $\frac{\partial J}{\partial \mathbf{a}^{(L)}}$ and $ \frac{\partial J}{\partial \boldsymbol{w}^{(L)}}$ using the chain rule~(\autoref{alg:clientHE}). 

\begin{align}
	\frac{\partial J}{\partial \mathbf{a}^{(L)}} &= \frac{\partial J}{\partial \hat{\mathbf{y}}} \frac{\partial \hat{\mathbf{y}}}{\partial \mathbf{a}^{(L)}}, \text{and} \ \ \ \
	\frac{\partial J}{\partial \boldsymbol{w}^{(L)}} &= \frac{\partial J}{\partial \mathbf{a}^{(L)}} \frac{\partial \mathbf{a}^{(L)}}{\partial \boldsymbol{w}^{(L)}}
 \end{align}

Following, the client sends $\frac{\partial J}{\partial \mathbf{a}^{(L)}}$ and $ \frac{\partial J}{\partial \boldsymbol{w}^{(L)}}$ to the server. Upon reception, the server computes $\frac{\partial J}{\partial \boldsymbol{b}}$ by simply doing $\frac{\partial J}{\partial \boldsymbol{b}}= \frac{\partial J}{\partial \mathbf{a}^{(L)}}$, based on equation~\eqref{eq:serverHELinear}. The server then updates the weights and biases of his linear layer according to equation~\eqref{equ:serverUpdateWB}.

\begin{align}
	\label{equ:serverUpdateWB}
	\boldsymbol{w}^{(L)} =  \boldsymbol{w}^{(L)} - \eta\frac{\partial J}{\partial \boldsymbol{w}^{(L)}}, \quad & b^{(L)} = \boldsymbol{b}^{(L)} - \eta\frac{\partial J}{\partial \boldsymbol{b}^{(L)}}
\end{align}

\noindent Next, the server calculates
\begin{equation}
	\frac{\partial J}{\partial \mathbf{a}^{(l)}} = \frac{\partial J}{\partial \mathbf{a}^{(L)}} \frac{\partial \mathbf{a}^{(L)}}{\partial \mathbf{a}^{(l)}},
\end{equation}
and sends $\frac{\partial J}{\partial \mathbf{a}^{(l)}}$ to the client. After receiving $\frac{\partial J}{\partial \mathbf{a}^{(l)}}$, the client calculates gradients of $J$ w.r.t the weights and biases of the Conv1D layer using the chain-rule: 
as:
\begin{align}
	\label{equ: gradients}
	\frac{\partial J}{\partial \boldsymbol{w}^{(i-1)}} &= \frac{\partial J}{\partial \boldsymbol{w}^{(i)}}\frac{\partial \boldsymbol{w}^{(i)}}{\partial \boldsymbol{w}^{(i-1)}} \\
	\frac{\partial J}{\partial \boldsymbol{b}^{(i-1)}} &= \frac{\partial J}{\partial \boldsymbol{b}^{(i)}}\frac{\partial \boldsymbol{b}^{(i)}}{\partial \boldsymbol{b}^{(i-1)}}    
\end{align}

After calculating the gradients $\frac{\partial J}{\partial \boldsymbol{w}^{(i)}}, \ \frac{\partial J}{\partial \boldsymbol{b}^{(i)}}$, the client updates $\boldsymbol{w}^{(i)}$ and $\boldsymbol{b}^{(i)}$ using the Adam optimization algorithm~\cite{kingma2015adam}.

\begin{algorithm}[!h]
\scriptsize 
\SetAlgoLined
 \textbf{Context Initialization:}\\
 $\displaystyle \ \ \ \ \ s.receive(\mathsf{ctx_{pub}})$\\
 \For{$\displaystyle \mathbf{ e} \ \text{in} \ \mathbf{E} $}{
 \For{$\displaystyle i \leftarrow 1 \ \mathbf{to} \ N \ $ }{
 $\displaystyle \mathbf{Forward\ propagation:}$\\
 $\displaystyle \ \ \ \ O.zero\_grad()  $\\
 $\displaystyle \ \ \ \ s.receive\ \overbar{(\mathbf{a}^{(l)})}$\\
 $\displaystyle \ \ \ \ \overbar{\mathbf{a}^{(L)}} \ \leftarrow \ \mathsf{HE.Eval} \left(f^{( i)}\left( \overbar{\mathbf{a}^{( l)}}\right)\right)$\\
 $\displaystyle \ \ \ \ s.send\left( \overbar{\mathbf{a}^{( L)}}\right)$\\
 $\displaystyle \mathbf{Backward\ propagation:}$\\
 $\displaystyle \ \ \ \ s.receive\left\{\frac{\partial J}{\partial \mathbf{a}^{(L)}} \& \frac{\partial J}{\partial \boldsymbol{w}^{(L)}}\right\}$\\
 $\displaystyle \ \ \ \ \text{Compute}\ \frac{\partial J}{\partial \boldsymbol{b}^{(L)}} $\\
 $\displaystyle \ \ \ \  \text{Update}\ \boldsymbol{w}^{(L)},\ \boldsymbol{b}^{(L)} $\\
 $\displaystyle \ \ \ \ \text{Compute}\ \frac{\partial J}{\partial \mathbf{a}^{(l)}} $\\
 $\displaystyle \ \ \ \ s.send \left( \frac{\partial J}{\partial \mathbf{a}^{(l)}} \right)$\\
 }
 }
 \caption{\textbf{Server Side}}
 \label{alg:serverHE}
\end{algorithm}

\section{Performance Analysis}
\label{sec:performance}

\subsection{ECG Datasets} 
\label{sebsec:ecgdataset}
We evaluate our method on 
the PTB-XL dataset~\cite{wagner2020ptb}.



\begin{figure}[!h]
	\centering
	\includegraphics[width=0.3\textwidth]{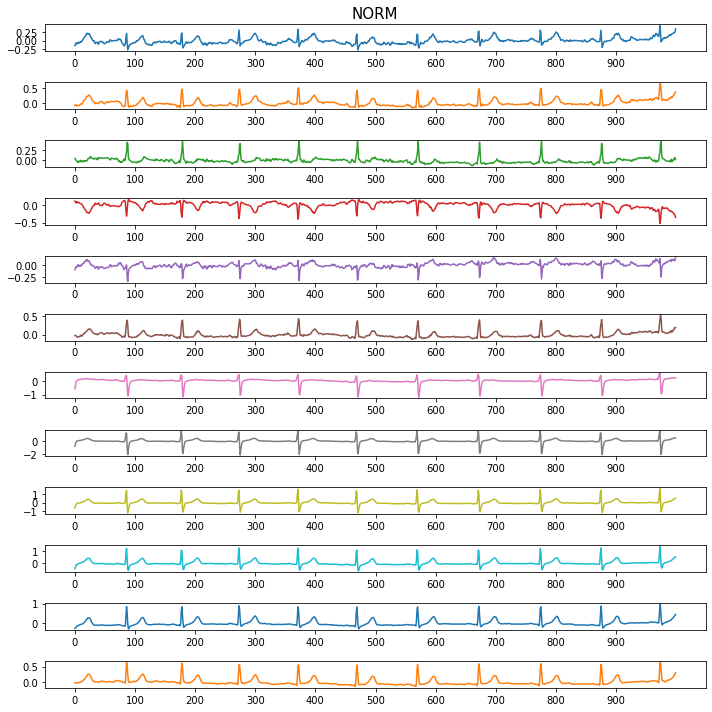}
	\caption{A 12-lead normal heartbeat from the PTB-XL dataset.}
	\label{fig:NORM-PTBXL}
\end{figure}

\paragraph{\textbf{PTB-XL:}}
According to~\cite{wagner2020ptb}, PTB-XL is the largest open-source ECG dataset since 
2020. The dataset contains 12-lead ECG-waveforms from 21837 records of 18885 patients. 
Each waveform from PTB-XL 
has a duration of 10 seconds. 
Two sampling rates 
are used to collect the data: 100 Hz and 500 Hz. In our experiment, we employ the 100 Hz waveforms. Each 12-lead ECG waveform is associated with one or several classes out of five classes: normal (NORM), conduction disturbance (CD), myocardial infarction (MI), hypertrophy (HYP), and ST/T change (STTC). For waveforms that belong to multiple classes, we choose only the first one and remove the others for simplicity. \autoref{fig:NORM-PTBXL} shows an example of a 12-lead normal heartbeat from the PTB-XL dataset. 
The dataset is then split into the train-test splits with 
a ratio of 90\%-10\%. In summary, after processing, we have a training split of size $[19267, 12, 1000]$, 
with 19,267 ECG waveform samples, 
of 12 channels (or leads) and 1,000 timesteps each. The test split's size is 
$[2163, 12, 1000]$.

\subsection{Experimental Setup}
\label{subsec:setup}
All neural networks are trained on a machine with Ubuntu 20.04 LTS, processor Intel Core i7-8700 CPU at 3.20GHz, 32Gb RAM, GPU GeForce GTX 1070 Ti with 8Gb of memory. We write our program in the~\href{https://www.python.org/downloads/release/python-397/}{Python programming language version 3.9.7}. The neural nets are constructed using the~\href{https://pytorch.org/get-started/previous-versions/}{PyTorch library version 1.8.1+cu102}. For HE algorithms, we employ the~\href{https://github.com/OpenMined/TenSEAL}{TenSeal library version 0.3.10}. 

In terms of hyperparameters, we train all networks with 10 epochs, a $\eta=0.001$ learning rate, and a $n=4$ training batch size. 
For the split neural network with HE activation maps, we use the Adam optimizer for the client model and mini-batch Gradient Descent for the server model. We use GPU for 
networks 
trained on the plaintext. For the U-shaped SL model on HE activation maps, we train the client model on GPU, and the server model on CPU.

\subsection{Evaluation}
\label{subsec:experiments}
In this section, we report the experimental results in terms of accuracy, training duration and communication throughput. 

The 1D CNN models used for the PTB-XL datasets have two Conv1D layers and one linear layer. The activation maps are the output of the last Conv1D layer. 


For the PTB-XL dataset~\footnote{For the MIT-BIH dataset, please have a look at~\cite{DBLP:conf/edbt/KhanNM23}}, the number of the input channels for the first Conv1D layer are 12, since the input data are 12-lead ECG signals. Besides, we only experiment with 8 output channels for the second Conv1D layer. We denote this network by $M$. Using $M$, the output activation map size is $[\text{batch size}, 2000]$.

\begin{figure}
	\centering
	\includegraphics[width=0.3\textwidth]{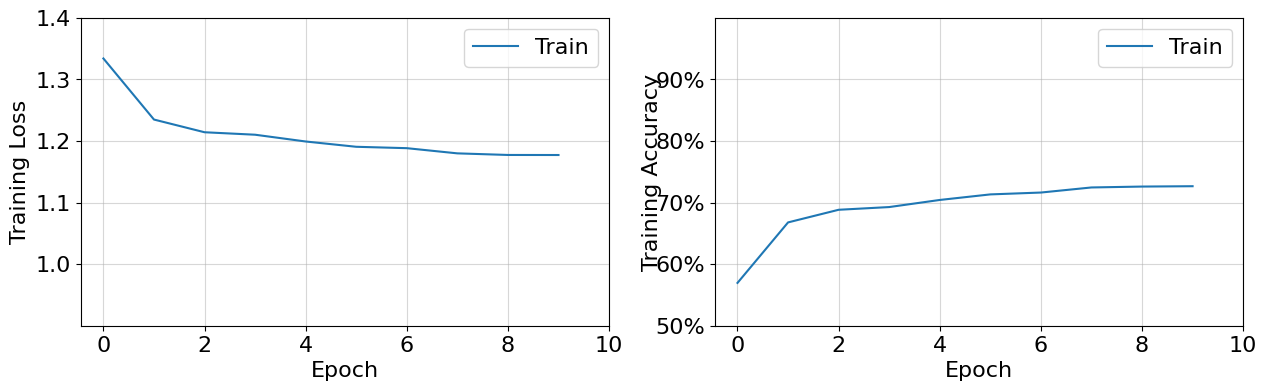}
	\caption{Training locally on the plaintext PTB-XL dataset.}
	\label{fig:localTrainPTBXL}
\end{figure}

Training the model $M$ locally (without any split) on the plaintext PTB-XL dataset results in \autoref{fig:localTrainPTBXL}. 
As we notice the best training accuracy after 10 epochs is only 72.65\%. The best test accuracy is 67.68\%. The reason for this low accuracy is that our neural network is small (it contains only 12013 trainable parameters), and we only train it for a 
particularly small epoch. Each training epoch takes 10.56 seconds on average.

\subsection{U-shaped Split Learning using Plaintext Activation Maps}
\label{subsec:plaintextactivationmap}
Based on our experiments, training U-shaped split model on plaintext yields same results in terms of accuracy compared to local training ~\cite{abuadbba2020can}. Although the authors of~\cite{abuadbba2020can} only used the vanilla version of the split model, they too found that, compared to training locally, accuracy was not reduced.

We will now discuss the training time and communication overhead of the U-shaped split models and compare them to their local versions. 
For $M$, the communication overhead is on average 316.9 Mb per epoch, which is much bigger due to the bigger activation maps sent from the client during training.


\subsection{Visual Invertibility}
\label{subsec:VisualInvert}

\begin{figure}[h]
	\centering
	\includegraphics[width=0.45\textwidth]{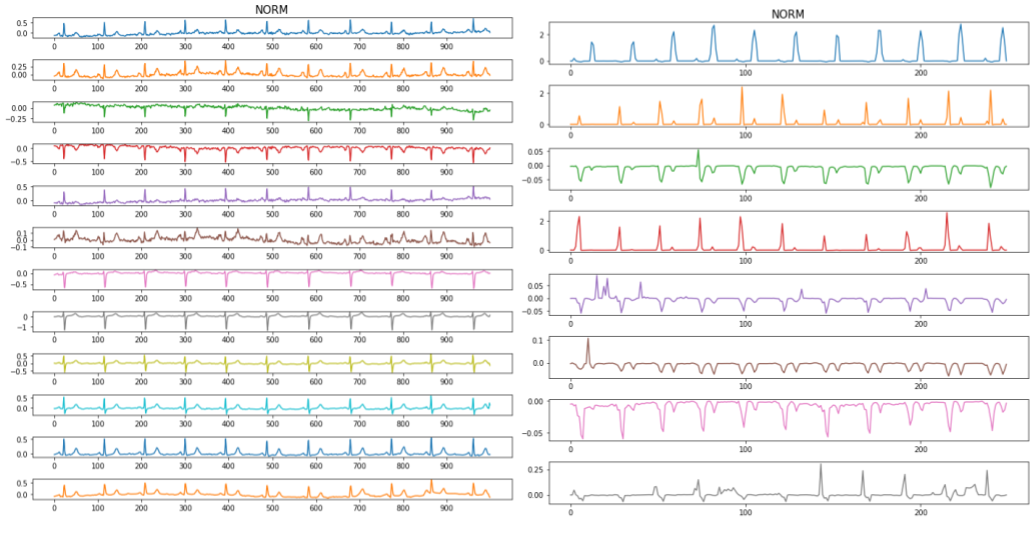}
	\caption{Visual invertibility of the model $M$ on the PTB-XL dataset. Left: input data (NORM class). Right: corresponding activation maps.}
	\label{fig:visual_invertibility_norm}
\end{figure}
\begin{figure}[h]
	\centering
	\includegraphics[width=0.45\textwidth]{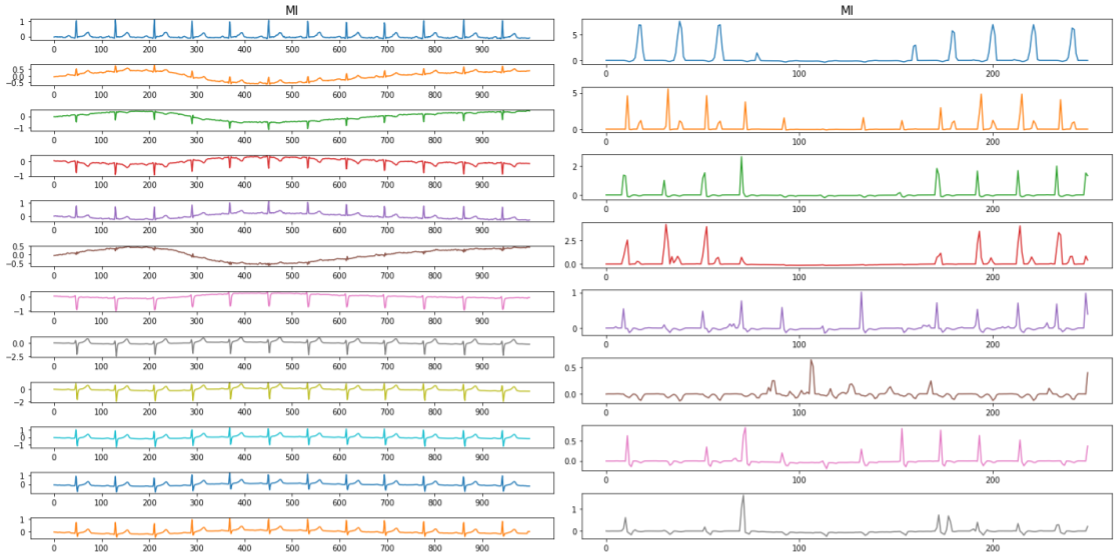}
	\caption{Visual invertibility of the model $M$ on the PTB-XL dataset. Left: input data (MI class). Right: corresponding activation maps.}
	\label{fig:visual_invertibility_mi}
\end{figure}
\begin{figure}[h]
	\centering
	\includegraphics[width=0.45\textwidth]{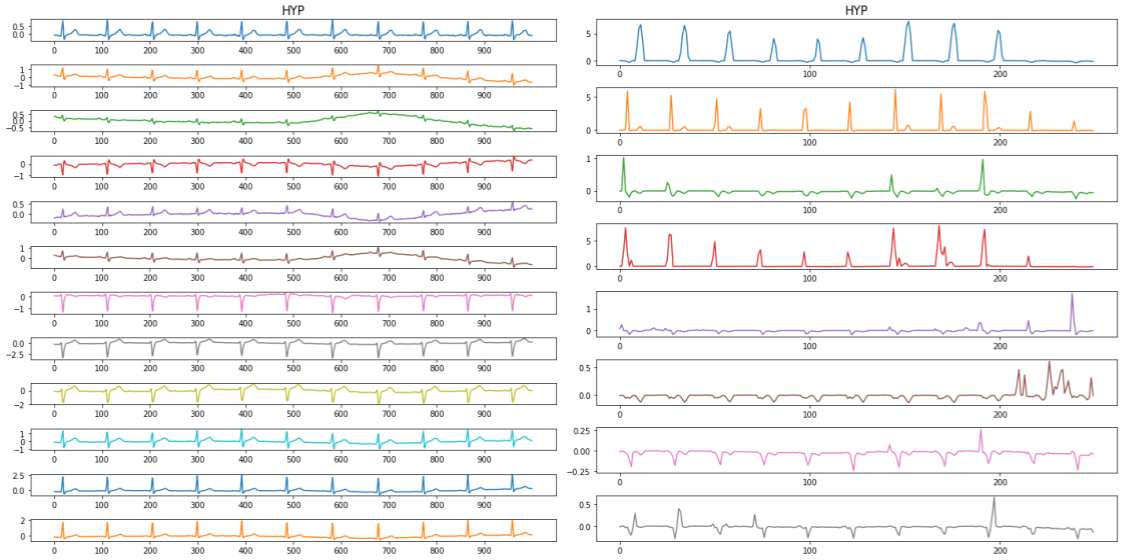}
	\caption{Visual invertibility of the model $M$ on the PTB-XL dataset. Left: input data (HYP class). Right: corresponding activation maps.}
	\label{fig:visual_invertibility_hyp}
\end{figure}
\begin{figure}[h]
	\centering
	\includegraphics[width=0.45\textwidth]{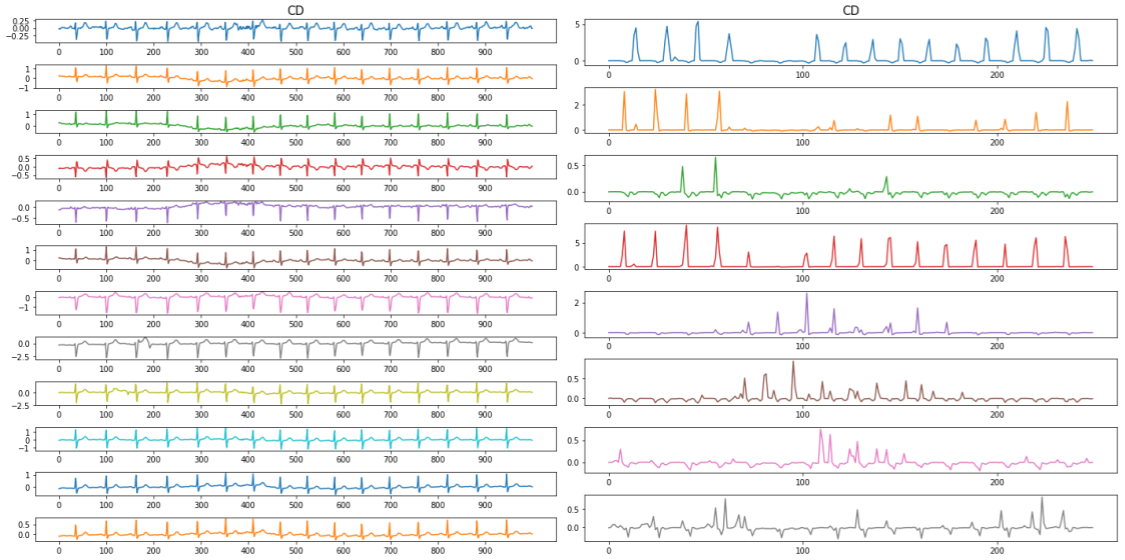}
	\caption{Visual invertibility of the model $M$ on the PTB-XL dataset. Left: input data (CD class). Right: corresponding activation maps.}
	\label{fig:visual_invertibility_cd}
\end{figure}
\begin{figure}[h]
	\centering
	\includegraphics[width=0.45\textwidth]{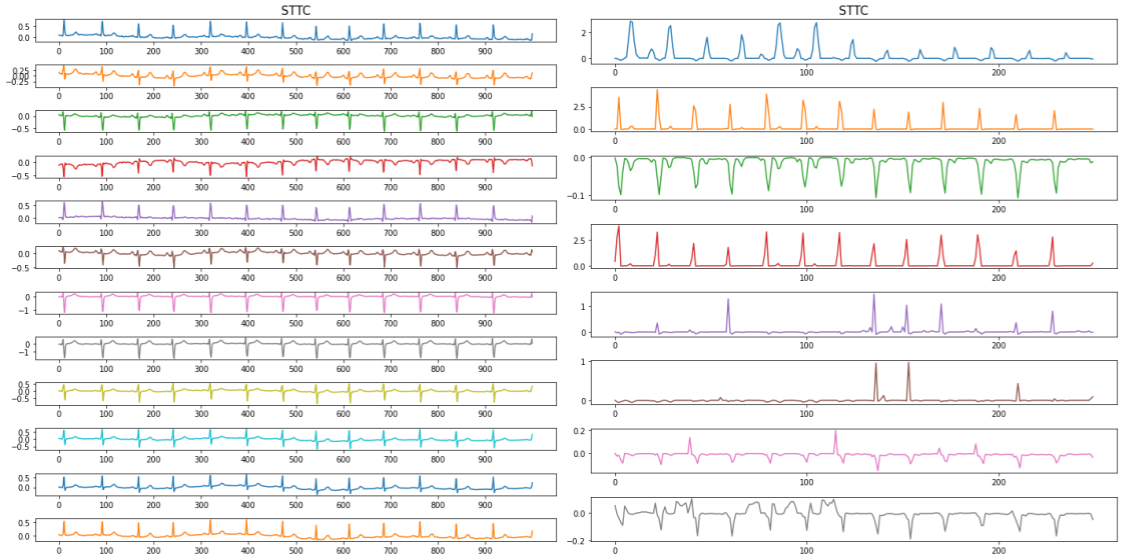}
	\caption{Visual invertibility of the model $M$ on the PTB-XL dataset. Left: input data (STTC class). Right: corresponding activation maps.}
	\label{fig:visual_invertibility_sttc}
\end{figure}

We visualize the activation maps produced by the model $M$ to access their visual similarity compared to the original signals. 
In SL, the activation maps are sent from client to server to continue the training process. 
A visual representation of the activation maps reveals 
a high similarity between certain activation maps and the input data from the client, as shown in \autoref{fig:visual_invertibility_norm}, \autoref{fig:visual_invertibility_mi}, \autoref{fig:visual_invertibility_hyp}, \autoref{fig:visual_invertibility_cd}, \autoref{fig:visual_invertibility_sttc} for five different classes of heartbeat in the PTB-XL dataset.

The figure indicates that, compared to the raw input data from the client, some activation maps have exceedingly similar patterns. This phenomenon clearly compromises the privacy of the client's raw data. The authors of~\cite{abuadbba2020can} quantify the privacy leakage by measuring the correlations between the activation maps and the raw input signal by using two metrics: distance correlation and Dynamic Time Warping. 
This approach allows them to measure whether their solutions 
mitigate privacy leakage work. Since our work uses HE 
, 
said metrics are unnecessary as  
the activation maps are encrypted.

\label{app:visual_invert_ptbxl}

\subsection{U-shaped Split 1D CNN with Encrypted Activation Maps}
\label{subsec:UShapedHE}

\begin{table*}
\centering
\caption{Training and testing results on the PTB-XL dataset. Training duration and communication are reported per epoch.}
\label{tab:trainingTestingResultsPTBXL}
\resizebox{0.7\textwidth}{!}{%
\begin{tabular}{c|c|l|c|c|c|c|c|c} 
\hline
\multirow{2}{*}{Network} & \multirow{2}{*}{Type of Network} & \multicolumn{4}{c|}{HE Parameters}                                   & \multirow{2}{*}{Training duration (s)} & \multirow{2}{*}{Test accuracy (\%)} & \multirow{2}{*}{Communication (Tb)}  \\ 
\cline{3-6}
                         &                                  & \multicolumn{1}{c|}{BE} & $\mathcal{P}$ & $\mathcal{C}$   & $\Delta$ &                                    &                                     &                                    \\ 
\hline
\multirow{6}{*}{$M$}   & Local                            & \multicolumn{4}{c|}{}                                                & 10.56                                  & 67.68                               & 0                                    \\ 
\cline{2-9}
                         & Split (plaintext)                & \multicolumn{4}{c|}{}                                                & 15.55                                  & 67.68                               & 316.9e-6                           \\ 
\cline{2-9}
                         & \multirow{4}{*}{Split (HE)}      & \multirow{3}{*}{True}   & 8192          & {[}40,21,21,40] & $2^{21}$   & 72 534                                 & 65.42                               & 115.64                              \\ 
\cline{4-9}
                         &                                  &                         & 4096          & {[}40,20,20]    & $2^{21}$   & 24 061                                 & 64.22                               & 18.20                               \\ 
\cline{4-9}
                         &                                  &                         & 4096          & {[}40,20,40]    & $2^{20}$   & 22 570                                 & 65.23                               & 18.77                               \\ 
\cline{4-9}
                         &                                  &                         & 2048          & {[}18,18,18]    & $2^{16}$   & 7 605                                  & 65.33                               & 1.93                                \\
\hline
\end{tabular}
}
\end{table*}


The results of training different settings of $M$ on the PTB-XL dataset are in \autoref{tab:trainingTestingResultsPTBXL}. 
The HE set of parameters $\mathcal{P}=8192$, $\mathcal{C}=[40, 21, 21, 40]$, $\Delta=2^{21}$ achieves the best test accuracy at 65.42\%. This result is only 2.26\% lower than the result obtained by the plaintext version. However, this set of parameters incurs the most communication overhead (115.64 Tb per epoch) and takes the longest to train (72 534 seconds per epoch). Overall, test accuracies achieved by different HE parameters are quite close to each other, with 65.42\% being the lowest. Interestingly, the smallest set of HE parameters $\mathcal{P}=2048, \mathcal{C}=[18, 18, 18], \Delta=2^{16}$ achieves the second-best accuracy at 65.33\% while only requiring 
about 1/10 of the training duration and 1/100 of the communication overhead compared to $\mathcal{P}=8192$. The two sets of parameters with $\mathcal{P}=4096$ produce quite similar results, 
while 
roughly taking same amount of time and communication overhead to train.

Through our experiments, we see that training on encrypted activation maps can produce very optimistic results, with accuracy dropping by 2-3\% for the best sets of HE parameters. Furthermore, training using BE can significantly reduce the amount of training time and communication overhead needed, while producing comparable results 
when it come to training without BE. The set of parameters with $\mathcal{P}=8192$ always achieve the highest test accuracy, though incurring the highest communication overhead and the longest training time. The set of parameters with $\mathcal{P}=4096$ can offer a good trade-off as they can produce on-par accuracy with $\mathcal{P}=8192$, while requiring 
significantly less communication and training time. Experimental results show that with the smallest set of HE parameters $\mathcal{P}=2048$, $\mathcal{C}=[18, 18, 18]$, $\Delta=2^{16}$, 
the least amount of communication and training time is required. In addition, this only works well when used together with BE. When training the network $M$ on the PTB-XL dataset, this set of parameters produces even better test accuracy compared to $\mathcal{P}=4096$. However, this result may be because the network $M$ is 
small. The test accuracy on the plaintext version is $67.68\%$, hence the noises produced by the HE algorithm 
do not yet have a significant role in reducing the model's accuracy. 

\noindent \textbf{Open Science and Reproducible Research:} 
To support open science and reproducible research, and provide researchers with the opportunity to use, test, and hopefully extend our work, the source code used for the evaluations is publicly available online\footnote{\href{https://github.com/khoaguin/HESplitNet}{https://github.com/khoaguin/HESplitNet}}.

\section*{Acknowledgment}
This work was funded by the HARPOCRATES EU research project (No. 101069535) and the Technology Innovation Institute (TII), UAE, for the project ARROWSMITH.

\section{Conclusion}
\label{sec:Conclusion}
In this paper, we focused on training ML models in a privacy-preserving way. We used the concept of SL in combination with HE and constructed protocols allowing a client to train a model in collaboration with a server without sharing valuable information about the raw data. To the best of our knowledge, this is the first work that uses SL on encrypted data. Our experiments show that our approach has achieved high accuracy, especially when compared with less secure approaches that combine SL with differential privacy. 

\balance
\bibliographystyle{ieeetr}
\bibliography{split}

\end{document}